\documentclass[useAMS,usenatbib]{mn2e}
\usepackage{times}
\usepackage{epsfig}

\usepackage{amsmath, amssymb, graphicx, multirow, rotating, lscape}

\begin{document}



\title{Multiwavelength Intraday Variability of the BL Lac S5~0716$+$714}



\author[Alok C. \ Gupta et al.]{Alok C.\ Gupta$^1$\thanks{Email: acgupta30@gmail.com},
T. P.\ Krichbaum$^2$,
P. J.\ Wiita$^3$, B.\ Rani$^{1,2}$, K. V.\ Sokolovsky$^{2,4,5}$,
\newauthor P.\ Mohan$^6$, A.\ Mangalam$^6$, N.\ Marchili$^2$, L.\ Fuhrmann$^2$,
I.\ Agudo$^{7,8}$, U.\ Bach$^2$, 
\newauthor R.\ Bachev$^9$, M.\ B{\"o}ttcher$^{10}$, K. E.\ Gabanyi$^{11,12}$, H.\ Gaur$^1$, K.\ Hawkins$^{10}$, G. N.\ Kimeridze$^{13}$, 
\newauthor O. M.\ Kurtanidze$^{13}$, S. O.\ Kurtanidze$^{13}$, C. -U.\ Lee$^{14}$, X.\ Liu$^{15}$, B.\ McBreen$^{16}$,
\newauthor R.\ Nesci$^{17}$, G.\ Nestoras$^2$, M. G.\ Nikolashvili$^{13}$, J. M.\ Ohlert$^{18,19}$, N.\ Palma$^{10}$, S.\ Peneva$^9$, 
\newauthor T.\ Pursimo$^{20}$, E.\ Semkov$^9$, A.\ Strigachev$^9$, J. R.\ Webb$^{21}$, H.\ Wiesemeyer$^{7}$, J. A.\ Zensus$^2$\\
\\
$^1$Aryabhatta Research Institute of Observational Sciences (ARIES), Manora Peak, Nainital, 263 129, India\\
$^2$Max-Planck-Institut f$\ddot{u}$r Radioastronomie (MPIfR), Auf dem H$\ddot{u}$gel 69, D-53121 Bonn, Germany\\
$^3$Department of Physics, The College of New Jersey, P.O.\ Box 7718, Ewing, NJ 08628, USA\\
$^4$Astro Space Center of Lebedev Physical Institute, Profsoyuznaya 84/32, 117997 Moscow, Russia \\
$^5$Sternberg Astronomical Institute, Moscow State University Universitetsky pr. 13, 119991 Moscow, Russia \\
$^6$Indian Institute of Astrophysics, Sarjapur Rd., Koramangala, Bangalore 560 034, India\\
$^7$Instituto de Astrof\'isica de Andaluc\'ia, CSIC, Apartado 3004, 18080, Granada, Spain \\
$^8$Institute for Astrophysical Research, Boston University, 725 Commonwealth Avenue, Boston, MA 02215, USA \\
$^9$Institute of Astronomy, Bulgarian Academy of Sciences, 72 Tsarigradsko Shosse Blvd., 1784 Sofia, Bulgaria\\
$^{10}$Astrophysical Institute, Department of Physics and Astronomy, Ohio University Athens, OH 45701, USA \\
$^{11}$F\"OMI Satellite Geodetic Observatory, PO Box 585, 1592 Budapest, Hungary \\
$^{12}$Konkoly Observatory, Research Centre for Astronomy and Earth Sciences, Hungarian Academy of Sciences, Konkoly Thege 
Miklos ut 15-17, Budapest, 1121, Hungary \\
$^{13}$Abastumani Observatory, Mt. Kanobili, 0301 Abastumani, Georgia \\
$^{14}$Korea Astronomy and Space Science Institute (KASI), Daejeon 305-348, Republic of Korea \\
$^{15}$Xinjiang Astronomical Observatory, the Chinese Academy of Sciences, 150 Science 1-Street, Urumqi 830011, China \\
$^{16}$UCD School of Physics, University College Dublin, Dublin, Ireland \\
$^{17}$Department of Physics, University La Sapienza, P.\ le Aldo Moro 2, I-00185, Roma, Italy \\
$^{18}$Astronomie Stiftung Trebur, Fichtenstrasse 7, 65468 Trebur, Germany \\
$^{19}$University of Applied Sciences, Wilhelm-Leuschner-Strasse 13, 61169 Friedberg, Germany \\
$^{20}$Nordic Optical Telescope, E-38700 Santa Cruz de La Palma, Santa Cruz de Tenerife, Spain \\
$^{21}$SARA Observatory, Florida International University, Miami, FL 33199, USA}
\date{Accepted 2012 June 19. Received 2012 June 15; in original form 2012 March 15}
\pagerange{\pageref{firstpage}--\pageref{lastpage}} \pubyear{2012}
\maketitle
\label{firstpage}

\begin{abstract}
We report results from a 1 week multi-wavelength campaign to monitor the BL\,Lac object
S5~0716$+$714 (on December 9-16, 2009). Nine ground-based telescopes at widely separated longitudes
and one space-based telescope aboard the {\it Swift} satellite collected optical data.
Radio data were obtained from the Effelsberg and Urumqi observatories,  X-ray data from 
{\it Swift}. In the radio bands the source shows rapid ($\sim (0.5-1.5)$ day) intra-day variability
with peak amplitudes of up to $\sim 10$\,\%. 
The variability at 2.8\,cm leads by about 1\,day  the variability
at 6\,cm and 11\,cm. This time lag and more rapid variations suggests an intrinsic contribution to  
the source's intraday variability at 2.8\,cm, while at 6\,cm and 11\,cm interstellar scintillation (ISS) 
seems to predominate. Large and quasi-sinusoidal variations of $\sim 0.8$ mag were 
detected in the V, R and I-bands. The X-ray data (0.2-10 keV) do not reveal 
significant variability on a 4 day time scale, favoring
reprocessed inverse-Compton over synchrotron radiation in this band.
The characteristic variability time scales in radio and optical bands are similar.
A quasi-periodic variation (QPO) 
of $0.9-1.1$\,days in the optical data may be present, but if so it is marginal 
and limited to 2.2 cycles. Cross-correlations between radio and optical
are discussed. The lack of a strong radio-optical correlation
indicates different physical causes of variability (ISS at long radio wavelengths, 
source intrinsic origin in the optical), and is consistent with a high jet opacity
and a compact synchrotron component peaking at $\simeq 100$\,GHz in an 
ongoing very prominent flux density outburst. For the campaign period, we construct a 
quasi-simultaneous spectral energy distribution (SED), including $\gamma-$ray data 
from the FERMI satellite. We obtain lower limits for the relativistic Doppler-boosting 
of $\delta \geq 12-26$, which for a BL\,Lac type object, is remarkably high.
\end{abstract}


\section{Introduction}

Blazars are the extreme subset of active galactic nuclei that are usually taken to include both 
BL Lacertae objects (BL Lacs) and flat spectrum radio quasars (FSRQs) because both
are characterized by rapid variability 
across the electromagnetic spectrum and  significant radio to optical polarization. Their spectra are 
generally well modelled by synchrotron and inverse Compton emission from relativistic jets aligned nearly 
($\lesssim$ 10$^{\circ}$) with the line of sight (e.g., Urry \& Padovani 1995). Blazar flux variations are 
seen on timescales ranging from a few  minutes through days and months to decades. Blazar variability 
timescales can be divided into three classes: changes from minutes to less than a day are variously
called microvariability, intra-night variability, or intra-day variability (IDV), which is the term we shall use; those from a 
few days to a few weeks are usually known as short timescale variability (STV);  flux changes from 
months to many years are called long term variability (LTV; e.g., Gupta et al.\ 2004). 

The vast majority of  optical  observations of blazars take place over single nights or several nights in 
succession at a given telescope, but these cannot provide the continuous coverage one would like to have in 
order to properly characterize and understand IDV and STV. The possibility that some coherent, even if 
temporary, fluctuations are present in blazar light curves is very important to investigate. Certainly one 
improves the chance of seeing such variations by having a long, densely sampled, light curve. These 
requirements have led to several intensive campaigns in the past that have looked at individual blazars 
with multiple telescopes at widely separated longitudes over a few days through several months
where simultaneous observations are obtained at different bands of the electromagnetic spectrum  
(e.g., Villata et al.\ 2000, 2006, 2008, 2009a, 2009b; Raiteri et al.\ 2003, 2005, 2007, 2008a, 2008b, 2008c; 
B{\"o}ttcher et al.\ 2005, 2009; Ostorero et al.\ 2006; Agudo et al.\ 2006; Fuhrmann et al.\ 2008;
Larionov et al.\ 2008, and references therein). 

We report results of a 7\,day long multi-telescope campaign on S5~0716$+$714 
with a focus on IDV in the optical and at some radio bands. 
The campaign was conducted on December 9 -- 16, 2009 
with the most dense sampling and best frequency coverage during December 11 -- 15.
There have been many studies of this blazar, mainly in optical bands, since it is bright, 
at a high declination and apparently always quite variable in the visible 
(e.g., Quirrenbach et al.\ 1991; Wagner et al.\ 1996; Sagar et al.\ 1999; 
Wu et al.\ 2005, 2007; Montagni et al.\ 2006; Stalin et al.\ 2006; 
Pollock et al.\ 2007; Gupta et al.\ 2008; Poon et al.\ 2009; Rani et al.\ 2010a, 2011; Chandra et al.\ 2011, 
and references therein). Our observing campaign also includes flux density measurements 
made in the centimeter and millimeter radio, optical and X-ray bands.

Theoretical models that seek to explain optical intra-day variability in AGN invoke several different 
mechanisms involving the accretion disk, including pulsation of the gravitational modes of the gaseous disk 
(e.g., Kato \& Fukue 1980; Nowak \& Wagoner 1992) or orbital signatures from ``hot-spots'' in the gas surrounding the black hole, either from the  disk itself or the corona above it (e.g., Zhang \& Bao 1991; Mangalam \& Wiita 1993). However, for blazars, particularly in high states, 
the variability almost certainly arises within the Doppler boosted relativistic jets and may well result 
from relativistic shocks in the jet (e.g.\ Marscher \& Gear 1985) or
turbulence behind such shocks (e.g.\ Marscher, Gear \& Travis 1992; Marscher et al.\ 2008),
from helical motion (e.g.\ Qian et al.\ 1991, Camenzind \& Krockenberger 1992), 
instabilities (e.g.\ Hardee et al.\ 2005), or slight changes in viewing angles (e.g.\ Gopal-Krishna \& Wiita 1992). 

A key motivation of this campaign was to look for possible correlations between radio and optical variability. A positive correlation could substantially constrain the origin of the variability in 
the radio bands and would strongly favor an intrinsic origin over one due to 
interstellar scintillation (e.g. Simonetti, Cordes \& Heeschen 1985, Rickett 1990). 
There have been earlier papers in which the variability properties of S5 0716+714 were discussed in both of these bands (e.g., Quirrenbach et al. 1991; Wagner \& Witzel 1995; Wagner et al.\ 1996; Ostorero et al.\ 2006;
Fuhrmann et al.\ 2008). 
Quirrenbach et al.\ (1991) reported correlated optical and radio variability in the source which, however, 
 was not again seen in later observational campaigns.  
Optical variability was claimed to be associated with changes in the radio 
spectral index (Qian et al.\ 1995 \& 1996).
This source is known to vary on different timescales (IDV to STV to LTV) and has been claimed to
exhibit quasi-periodic oscillations (QPOs) on all these timescales. 
On IDV timescales  0716$+$714 has shown QPOs on various occasions 
with timescales ranging from $\sim$ 15 min to 73 min (Gupta et al.\ 2009; Rani et al.\ 2010b). 
On STV timescales there is weak evidence for quasi-periodicity appearing on 
timescales of $\sim$ 1 day and $\sim$ 7 days (Quirrenbach et al.\ 1991; 
Wagner 1992; Heidt \& Wagner 1996). On LTV timescales claims have been made 
for possible near-periodicities of $\sim$ 3.0$\pm$0.3 years (Gupta et al.\ 2008).   
We adopt a redshift of $z = 0.31$ for this source (Nilsson et al.\ 2008) but the lack of spectral lines means that it is still quite uncertain ($\pm 0.08$).

In \S 2 we describe the observations and data reductions.  We present our results in \S 3 and
  \S 4 includes a discussion and our conclusions.

\section{Observations and Data Reductions}

\subsection{Radio data}

The source S5\ 0716+714 was observed on December 11--15, 2009 with the MPIfR 100m telescope at Effelsberg, Germany and with the 25m Nanshan radio telescope of 
the Urumqi Observatory, China. The observations and source selection at both observatories
were coordinated and done in a similar manner. At Effelsberg the observations were performed at 3 
frequencies using the secondary focus heterodyne receivers operating at 2.7, 4.85 and 10.5 GHz.
At Urumqi the observations were done using a Cassegrain focus receiver, which operates
at a frequency of 4.8~GHz. At both telescopes the flux density 
measurements were made in a similar way, with repeated cross-scans in azimuth and 
elevation. Frequent switching between target and calibrator sources on time scales of a few minutes 
allowed us to monitor the gain variations introduced by the receiving system and the atmosphere. The 
gain variations seen in the calibrator sources were used to improve the flux density calibration and 
correct for time and elevation dependencies. This observing and calibration technique is well established
and has been applied before for various intra-day variability observations. The
details of the observing strategy, the receiver parameters and the calibration methods
have been described before (Gabanyi et al.\ 2007; Fuhrmann et al.\ 2008; Marchili et al.\ 2010). 
As secondary calibrators the two non-variable radio sources 0951+699 and 0836+710 were 
used. The absolute flux density scale was set using 3C 286 and NGC 7027, adopting 
the scale of Zijlstra et al. (2008). 

\subsection{Optical data}

Our official campaign on S5~0716$+$714 began on 11 December 2009 and ran through 15 December 2009.
Additional data was obtained on 9, 10, 16 and 17 December, though the coverage was not as dense.
We briefly describe the telescopes and cameras that were involved in these observations.  We note that as
this source is point-like the different apertures used at different telescopes have negligible effects on the
measured fluxes and agreement was excellent whenever different telescopes provided overlapping data.

The source was observed on 12 December 2009 with the 23.5 cm f/10 Schmidt-Cassegrain telescope located
on the roof of the Department of Physics of La Sapienza University (Roma, Italy), with a Bessel
R filter and a CCD camera with a Peltier cooled Kodak KAF 1603ME CCD chip. Due to the lack of a guiding
system, images were taken in sequences of 9-10 frames with 60 s exposures and then stacked and
summed together to increase the S/N ratio. Sky flats were used for flat fielding. Data reduction
and analysis were performed with standard Image Reduction and Analysis Facility \texttt{(IRAF)}\footnote{IRAF
is distributed by the National Optical Astronomy Observatories, which are operated by the Association
of Universities for Research in Astronomy, Inc., under cooperative agreement with the National Science
Foundation.} routines. Aperture photometry was made with a 3.3 arcsec (2 pixels) radius. Simultaneous
observations were made with the 31 cm f/4.5 Newtonian telescope located at Greve in Chianti (near
Florence, Italy) equipped with a CCD camera with a back-illuminated Peltier cooled  Site SIA502A chip
and B, V, R, I Bessel filters. A four color sequence was made at the beginning of observations and then R
filter monitoring began with 180 s exposure times. \texttt{IRAF} standard tasks were used for data reduction and
analysis; aperture photometry was made with a 7 arcsec (2 pixels) radius. Photometric errors were
estimated from the rms deviations of the reference stars, and found to be generally less than 0.01 mag.
The light curves of the two instruments were in very good agreement. Reference standard stars were taken
from Villata et al.\ (1998) in the same blazar field.

At the MDM Observatory on the south-west ridge of Kitt Peak, Arizona, USA, data were taken for limited
periods during the nights of 9, 10, 11 and 12 December with the 1.3 m McGraw-Hill Telescope, using the
Templeton CCD with B, V, R, and I filters. The standard data reduction, using \texttt{IRAF}, included bias
subtraction and flat-field division. Instrumental magnitudes of S5~0716$+$714 plus four comparison stars
in the field (Villata et al.\ 1998) were extracted using the \texttt{IRAF} package \texttt{DAOPHOT}\footnote{Dominion
Astrophysical Observatory Photometry software} (Stetson 1987) with an aperture radius of 6 arcsec and
a sky annulus between 7.5 and 10 arcsec.

The observations at the Abastumani Observatory were conducted on 9, 11, 12, 14, 15 and 16 December 2009
at the 70-cm meniscus telescope (f/3). These measurements were made with an Apogee CCD camera Ap6E
(1K $\times$ 1K, 24 micron square pixels) through a Cousins R filter with exposures of 60--120 sec.
Reduction of the image frames were done using \texttt{DAOPHOT II}. An aperture radius of 5 arcsec was used
for data analysis.

Observations of S5~0716$+$714 were carried out on 13 December  2009 using the 50/70-cm Schmidt
telescope at Rozhen National Astronomical Observatory, Bulgaria. The telescope is equipped with a
FLI (Finger Lakes Instrumentation) ProLine PL16803 CCD (4096 pixels $\times$ 4096 pixels), BV Johnson and RI Cousins filters.
Standard data reduction used \texttt{MIDAS} and included bias subtraction and flat-field division.
Instrumental magnitudes of S5 0716+714 and comparison stars in the field (Villata et al.\ 1998)
were extracted using the \texttt{MIDAS} package \texttt{DAOPHOT} with an aperture radius of 5.6 arcsec (2$\times$FWHM).

Observations of the source on 11, 13 and 15 December 2009 were carried out using the 1.04 m Sampurnanand
telescope (ST) located at  Nainital, India. It has  Ritchey-Chretien optics with a f$/$13 beam equipped
with a CCD detector which is a cryogenically cooled 2048 pixel $\times$ 2048 pixel chip mounted at the
Cassegrain focus and Johnson UBV and Cousins RI filters.
Each pixel of the CCD chip has a dimension of 24 $\mu$m$^{2}$,
corresponding to 0.37arcsec$^{2}$ on the sky, thereby covering a total field of $\sim$
13$^{\prime}$ $\times$ 13$^{\prime}$.
Data processing (bias correction, flat-fielding and cosmic ray removal) was done by the standard routines
in \texttt{IRAF} and photometry of the blazar and standard stars in its field employed a stand-alone version of
\texttt{DAOPHOT II}.  Aperture photometry was carried out with four concentric  aperture radii, i.e.,
$\sim$ 1$\times$FWHM, 2$\times$FWHM, 3$\times$FWHM and 4$\times$FWHM. We found that aperture
radii of 2$\times$FWHM almost always provided the best S/N, so we adopted that aperture for our final results.

Observations  also were carried out on 13 December 2009 using the 61-cm Boller and Chivens
reflector at Sobaeksan Optical Astronomy Observatory in Korea. The FLI CCD camera with thermo-electric
cooled Kodak KAF 4301 2K CCD chip set and standard R Cousins filter were used for the observations.
All images were reduced using standard \texttt{IRAF} tools. Aperture photometry parameters that maximized the S/N
were an aperture radius of 1.5 $\times$ FWHM, an inner radius of the sky annulus of 5 $\times$ FWHM, and a
sky annulus width of 10 pixels.

On 12, 13 and 14 December 2009 the source was observed at the 1.2 m Cassegrain telescope at the Michael Adrian
Observatorium of Astronomie Stiftung Trebur, Germany.  A Roper Scientific EEV 1340 EB CCD camera and a Cousins
R filter were employed. The data were reduced using \texttt{MIRA Pro 7} software. An aperture radius of 2.9 arcsec,
and inner and outer sky annulus radii of 10.4 and 13.8 arcsec, respectively, were used.

Observations of S5~0716$+$714 were carried out on 12 December 2009 at the  0.9 m optical SARA (Southeastern Association 
for Research in Astronomy) telescope
at Kitt Peak National Observatory, USA. The camera was an Apogee U42 with Johnson and Cousins UBVRI filter
set. \texttt{MIRA} software was used for image processing and data analysis. The data were analyzed with aperture size of 4 arcsec. The  blazar light curve was calibrated using the local standard stars present in the blazar field (Villata et al.\ 1998).

The final set  of ground based optical observations of S5~0716$+$714 were carried out on 10 and 16 December 2009 at the 2.56 m Nordic Optical 
Telescope (NOT), Canary Islands, Spain using ALOFSC (10 Dec) and MOSCA (16 Dec) using UBVRI filters. The data 
were reduced using standard \texttt{IRAF} procedures, including those for debiasing and twilight flat-correction. The photometry was done using  
\texttt{IRAF/APPHOT} with the aperture radius chosen to be close to the FWHM of the image.

S5~0716$+$714 also was observed by the {\em Swift} satellite's Ultraviolet-Optical Telescope (UVOT, Roming et al.\ 2005)
in the V-band during 54 pointings conducted 
between December 11 and 15, 2009.
The pointings were typically
separated by the 96 minutes period of {\em Swift's} orbital revolution. Observations from
the space-based platform have the obvious advantage of not being interrupted by the day/night cycle or
unfavorable weather, and therefore can probe variability on timescales of hours to days with dense and
uniform sampling. The UVOT is a D = 300~mm, f = 3810~mm modified Ritchey-Chretien telescope equipped with
a micro-channel plate intensified CCD detector operated in photon counting mode. Detectors of this type may
provide information about the time of arrival of individual photons and they have no specific saturation limit,
but these advantages come at the cost of a nonlinear response to the number of incoming photons because of the dead time after
each registered photon event known as the coincidence loss (or pile-up) effect (Poole et al.\ 2008, Breeveld et
al.\ 2010). The \texttt{VaST} software (Sokolovsky and Lebedev 2005) based on the SExtractor code (Bertin and Arnouts
1996) has been applied to conduct aperture photometry of the UVOT images. The \texttt{VaST} software is designed to
deal with imaging data obtained with non-linear detectors (e.g. Kolesnikova et al.\ 2008, 2010) and has been
successfully applied for the UVOT data reduction before (Sokolovsky 2009). The magnitude scale was set using
comparison stars 4-8 from Villata et al.\ (1998). The  data reduction technique we employed allows us to avoid
uncertainties in the coincidence loss correction for bright sources and enables direct comparison of the UVOT
results with those obtained with ground-based telescopes after a R-V color correction of $0.43 \pm 0.04$ mag is
applied (Rani et al.\ 2010a).

\subsection{X-ray data}

{\it Swift}'s X-ray telescope (XRT, Burrows et al. 2005) 
observed the source simultaneously with the UVOT in the $0.2$--$10$~keV
energy range. The XRT was operating in the photon counting (pc) mode.
The total exposure of 15.5~ksec was collected during 54 satellite
pointings performed between 2009 December 11--15;
each pointing was about 300~sec long.
Observations were processed using the \texttt{xrtpipeline} tool
from the \texttt{HEASOFT v6.8} package applying the standard filtering criteria.
The mean count rate during the observations was $0.53\pm0.01$~cts/sec
which allows us to neglect the pile-up effect.

For the spectral analysis, all sub-exposures were combined into a single
event file using \texttt{XSELECT}, the corresponding exposure maps were
generated with \texttt{xrtexpomap} and combined using \texttt{XIMAGE}.
The source spectrum was extracted from circular region of 20~pixel radius.
The background counts were extracted from a region away from the source.
The auxiliary response file (ARF) was generated with \texttt{xrtmkarf},
and the response matrix \texttt{swxpc0to12s6$\_$20070901v011.rmf}
was used.
The spectrum was re-binned with the tool \texttt{grppha} to
contain at least 25 counts per energy bin to enable use of the $\chi^2$
statistic. It was modeled with an absorbed power law, taking the
neutral hydrogen column density 
fixed to the Galactic value of $N_\mathrm{HI} = 2.0 \times 10^{20}$~cm$^{-2}$ from
Kalberla et al.\ (2005) while the photon index, $\Gamma$, and the normalization
were left as free parameters. The spectral modeling was conducted in
\texttt{XSPEC v12.5.1n}.

For the light curve analysis, the same procedure was applied to
each individual sub-exposure.  The value of $\Gamma$ was fixed to that
obtained from the combined spectrum to decrease the uncertainty of the
flux measurements. Unabsorbed fluxes were computed by integrating
the power law model in the $0.2$--$10$~keV energy range.

\section{Analysis and Results}

\subsection{Variability curves: general trends and behavior}

\subsubsection{Radio} 

We characterize the variability parameters in the same way as in
Fuhrmann et al.\ (2008), by the variability index, $m$, the noise-bias corrected
variability amplitude, $Y$, and a reduced chi-square value, $\chi^2_r$, for a fit to a 
constant flux\footnote{The source is considered to be variable if the $\chi^2$-test gives a probability
of $\leq 0.001$  for the assumption of constant flux density}. 
In Table 1 we summarize the results. In column 1 we give the 
observing frequency, in column 2 the mean flux density and  its error.
In column 3 the variability index
of 0716$+$714 is given, while the next column
contains the variability index of the secondary calibrators (0951+699 and 0836+710), which
is a measure of the residual calibration errors.
Column 5 has the variability amplitude $Y$, followed by the reduced $\chi^2_r$ in column 6, the number
of measurements in column 7, the reduced $\chi^2_r$ value corresponding to a significance level of 99.9 per cent
in column 8 and the observing telescope in the last column. 

\begin{table*}
\caption{ The variability analysis parameters at radio wavelengths }
\begin{tabular}{llcccclcl} \hline
Freq.       &   Mean Flux     &  m  &$m_0$&   Y  &$\chi^2_r$&  N   &$\chi^2_{r~99.9\%}$& Telescope  \\       
GHz (cm)    &      (Jy)       &(\%) & (\%)& (\%) &          &      &               &            \\\hline
2.7 (11.0)  &1.03$\pm$0.024   &2.35 &0.30 & 7.0  & 63.595   &127   & 1.44          &Effelsberg  \\
4.85 (6.0)  &1.46$\pm$0.024   &1.67 &0.25 & 4.94 & 24.360   &132   & 1.44          &Effelsberg  \\
4.85 (6.0)  &1.46$\pm$0.029   &2.01 &0.33 & 5.94 & 16.256   &117   & 1.45          &Urumqi       \\
10.5 (2.8)  &2.45$\pm$0.048   &1.96 &1.10 & 4.87 & 4.865    &101   & 1.49          &Effelsberg   \\\hline
\end{tabular}  \\
m=variability index = $\sigma_{S}$/$<$S$>$,$\sigma_{S}$ standard deviation   \\
$m_0$ = variability index of the secondary calibrators \\
Y= 3$\sqrt{m^2 - m_0^2}$ = bias corrected variability amplitude (see Fuhrmann et al. 2008) \\
$\chi^2$= reduced Chi-square \\
N= number of data points \\
$\chi^2_{r~99.9\%}$ = reduced Chi-square corresponding to a significance level of 99.9$\%$ \\
\end{table*}

As it is seen from Fig.\ \ref{fig1}, the overall variability is mild, with
a decreasing trend of $\delta S /<S>$ of $\sim 8$ per cent at 2.7 and 4.8~GHz, and a $\sim$ 7 per cent
increase in $\delta S /<S> $ toward the end of the experiment at 10.5 GHz. 
At all frequencies the source varies on time scales of 0.5--1 days
with a variability index $m$ of about 1--2 per cent. Since the influence of the atmosphere
and residual calibration errors increase with frequency, we also see a slight increase
of the variability index of the secondary calibrators ($m_0$) with increasing frequency, 
from 0.25--0.33 per cent at 2.7 GHz and 4.8~GHz to 1.1 per cent at 10.5 GHz. 
The calibration-bias corrected variability index $Y$ takes into account the
calibration uncertainties in $m_0$ and their frequency dependence. 
We therefore see a decrease of the variability amplitude $Y$ for 0716+714 from 
7 per cent at 2.7 GHz to 4.9 per cent at 10.5 GHz. Thus the rms-amplitude of the intra-day 
variability of 0716$+$714 decreases with increasing frequency, as it could be 
expected for (weak) refractive interstellar scintillation (e.g.\ Walker 1998, Beckert et al.\ 2002).
We however note that our present finding is opposite to an observed increase 
of the variability amplitude with frequency for inter-day variability in this source when it was
observed in November 2003 (Fuhrmann et al.\ 2008). Temporal changes of the
frequency dependence of the RADIO variability index in 0716+714 are well known and
can be interpreted as a result of variable source size and the related change
of the relative dominance of interstellar scintillation over source intrinsic
variability (cf.\ Krichbaum et al.\ 2002).

We qualitatively compare the observed modulation indices 
with those obtained from an analytical solution for the 
strength of interstellar scintillation (ISS) in the weak, 
quenched scattering regime (using equations of Beckert et al.\ 2002; see 
also Fuhrmann et al.\ 2008). For a plasma screen of 1\,pc thickness 
and a turbulence strength similar to the value obtained from pulsar 
scintillation in the Local Bubble, the observed variability indices shown in Table 1
can be reproduced by assuming a scintillating component of size 
$\theta \sim 0.1-0.2$\,mas (VLBI core size at 5\,GHz, Bach et al.\ 2006) and an adopted 
screen distances of $\sim 50-200$\,pc for the Local Bubble (Bhat et al. 1998).
We further assume that the scintillating source component 
contains about 70\, per cent of the source's total flux density 
as indicated from the VLBI compactness and re-scale the 
observed modulation indices accordingly. Figure \ref{fig2}
shows the scaled and noise bias corrected variability index ($\sim \sqrt{(m/0.7)^2 - m_0^2}$)
plotted versus frequency together with models for  screens
at 100\,pc and 200\,pc distances. The data are in good agreement 
with this ISS slab model, which also reproduces the observed variability
time scale of the order of $\sim 1$\,day (see following sections). 
At 10.5\,GHz, however, the observed variability index is larger than expected 
from the model. At 4.8 GHz an annual modulation of the intra-day
variability time scale has been found which is caused by ISS (Liu et al. 2012). 
Underlying source intrinsic variability may explain this excess. 
If we use only the first 2 days of the 2.8\,cm data and by this omit 
the flux density rise seen at the end of the observations 
(see Fig. \ref{fig1}), the resulting lower variability 
index would agree with the expected frequency dependence (open and filled
symbols in Fig. \ref{fig2}).

\noindent
\subsubsection{Optical}
\begin{figure*}
   \centering
\epsfig{figure = 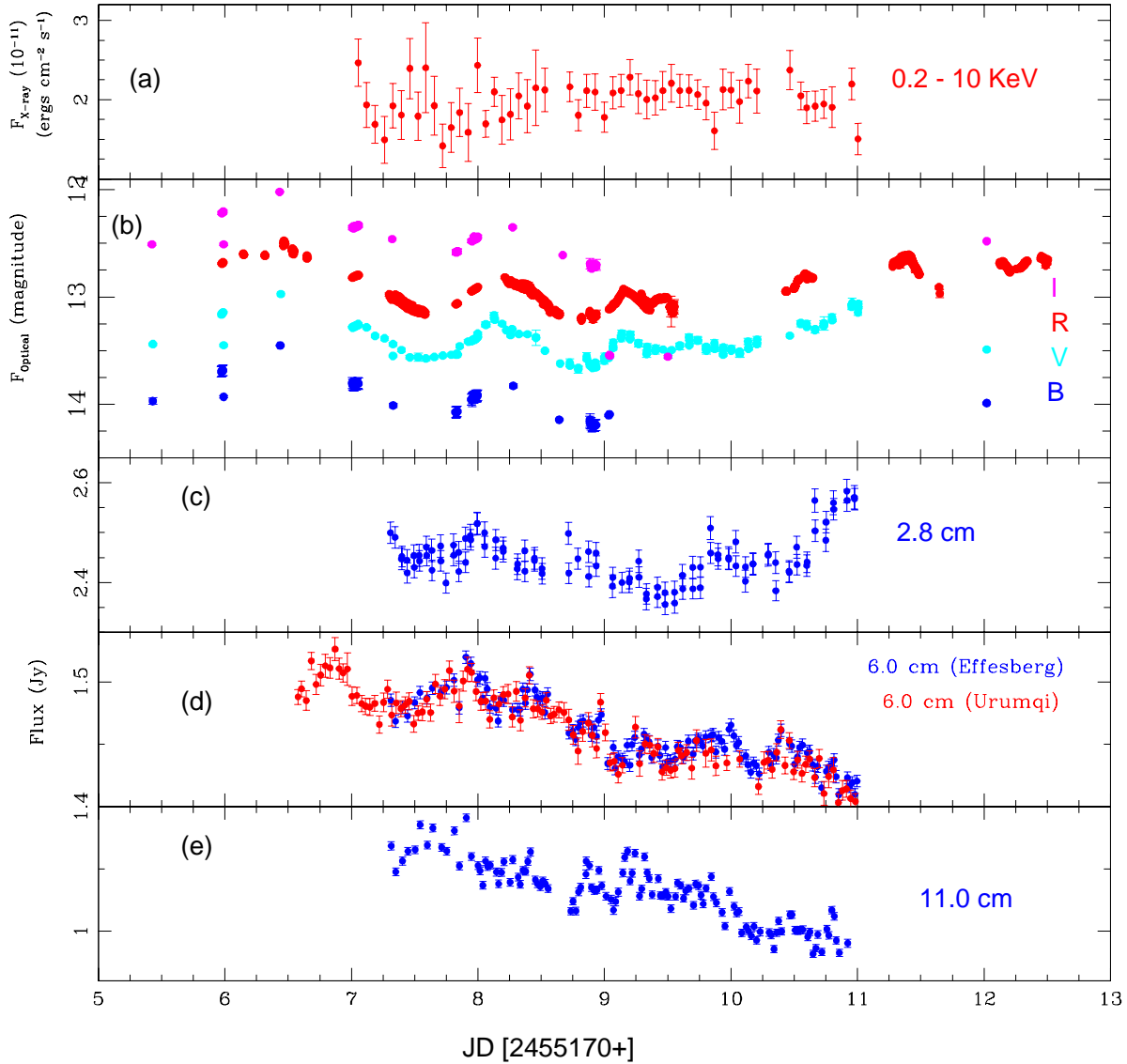,scale=0.8}
\caption{The broad band (radio through optical to X-ray ) variability observed in the 
source S5 0716+714 over the campaign period. (a) 0.2$-$10 KeV light curve of the source; 
(b) optical light curves of the source in B,V,R and I passbands; 
(c) flux density variability at 2.8 cm wavelengths; (d) the flux density variability at 6.0 cm 
from two different telescopes (Effelsberg and Urumqi) and (e) flux density variability at 11.0 cm
wavelengths.}
\label{fig1}
\end{figure*}

\begin{figure}
\epsfig{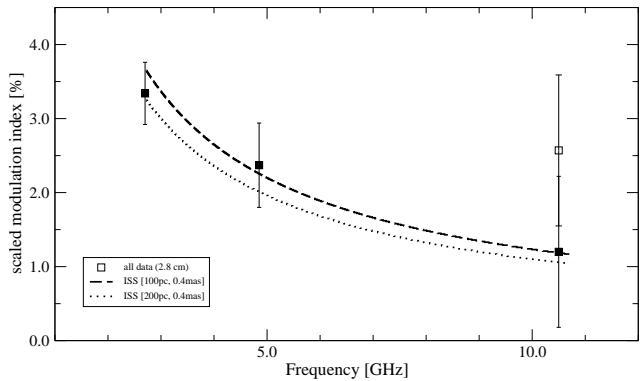}
\caption{Variability index plotted versus frequency. At 10.5\,GHz (2.8\,cm) 
the open symbol is for the whole data set and the filled square represents the truncated 
data set without the flux density increase seen at the end (see text). The lines show 
a model for a thin scattering screen located at a distance of 100\,pc (dashed line)
and 200\,pc (dotted line). }
\label{fig2}
\vspace*{0.5cm}
\end{figure}

The variability is even more pronounced at optical frequencies. The optical 
light curves of the source in the  B, V, R and I passbands shown in Fig. \ref{fig1}(b) display 
simultaneous flaring trends.   Instead of a continuous rise or decay, multiple peaks and troughs
of the flux were observed that give the appearance that the source might be
showing  nearly periodic 
variations. Although the B and I band light curves are less densely sampled 
compared to V and R, this  behavior is consistent through all the 
optical passbands.  The variability timescales are also
comparable at all four optical passbands.  Detailed discussions 
of  the variability timescales and a search for possible quasi-periodic oscillations (QPOs)
in the R-band light curve are given in Sections 3.2 and 3.6, respectively.

\noindent
\subsubsection{X-ray} 
While the source was highly active at optical frequencies no significant variation of 
X-ray flux was detected during the campaign (see Fig. 1(a)). The $\chi^2_r$ test returns 
a 0.75 probability that the 
entire observed light curve is the result of random noise.
From the visual inspection of the light curve it seems that in the beginning
of the campaign (before JD2455178.6) the source exhibits some variation but it
remains quiet during the second half of the campaign. The Kolmogorov-Smirnov (K-S)
test gives only a 0.016 probability that flux measurements obtained before JD=2455178.6
and after this date are drawn from the same parent distribution. The $\chi^2_r$ test
returns a 0.58 probability that values before JD2455178.6 are just random
noise while this rises to a 0.85 probability for the values after JD=2455178.6.
It is possible that the object exhibits small intraday variability in X-rays, but we were 
not able to detect it significantly with these observations; the error bars are just too large.

\subsection{Variability time scales}

To estimate any time scales present in the variability at optical and radio wavelengths we 
first employed the 
structure function (SF) analysis method. We followed Rani et al.\ (2009) to calculate the 
structure functions of the observed radio and optical light curves. 
Figure \ref{fig3} shows the SF for the  three radio light curves at 11, 6 and 2.8 cm  
wavelengths. At 11 cm and 6 cm the amplitude and time 
scale of the variability are very similar, showing a break in the structure function (indicating 
a characteristic variability time scale) at $t_{min} = 1.5\pm0.1$ day at 11\,cm and
$t_{min} =1.1 \pm 0.1$ day at 6\,cm. The latter 
time scale is visible in both, the 6\,cm data sets from Effelsberg and Urumqi.
At 11\,cm and 6\,cm another break near $t_{max} = 2.9\pm0.1$ days is 
only marginally significant, since it is longer than
half of the duration of the observations. For this slower variability ($t > 2.5-3$ days) 
the data indicate higher variability amplitudes at 11 cm than at 6 cm. The SF curves at 11 cm and 
6 cm wavelengths are characterized by two different slopes. The SF curve at 11 cm follows a  
power-law slope, $\beta = 0.90\pm0.04$ until $t_{min}$ (with $SF \propto t^{\beta}$) 
and then $\beta = 2.02\pm0.11$ from $t_{min}$ to 
$t_{max}$, while at 6 cm $\beta = 0.98\pm0.05$ until $t_{min}$ and $\beta = 1.67\pm0.06$ from 
$t_{min}$ to $t_{max}$.

The  SF of the data at 2.8 cm, however, shows a more pronounced and faster 
variability than for the two longer wavelengths. This is also seen in the auto-correlation
functions (see Section 3.3 and Fig. \ref{fig6}). The characteristic time scales of variability at 
2.8 cm wavelength are $t_{min} = 0.50\pm0.05$ days and $t_{min} = 1.4\pm0.1$ days and the 
SF curve is characterized by a slope $\beta = 0.65\pm0.06$ until $t_{min}$ and $\beta = 
1.37\pm0.17$ from $t_{min}$ to $t_{max}$. 
Below $t  \approx 3$ days, the variability at 2.8 cm appears more pronounced
than that at 11 cm and 6 cm. However, Table 1 shows that the measurement
uncertainty $m_0$ at 2.8 cm is 1.1 per cent, which is 3 to 4 times higher than at 11 cm or 6 cm. 
Although the formal significance of variability at 2.8 cm is higher than 99.9 per cent, 
the larger measurement uncertainty at this wavelength should lead to a cautious interpretation
of the variability in this band. While in the first half of the 2.8 cm light curve
the variability is very low ($m=1.15$ per cent), we note in the second half a continuous rising 
trend in flux density, which is the main reason for the high significance of the overall variability. Since
a departure from this rising trend is not seen, the true variability amplitude could 
be higher and the variability time scale could be longer than 1.4 days. 

In summary we find that the fastest and most pronounced variability is seen
at the shortest radio wavelength (2.8\,cm) with a characteristic time scale of 
$0.5\pm0.05$ days. Common to the three radio data sets (11, 6, and 2.8\,cm) 
is also a slower variability mode, which appears at a characteristic timescale 
of 1.1 days at 6\,cm and to 1.4-1.5 days at 11\,cm and 2.8\,cm. 
The structure function at 2.8\,cm differs substantially from the 
more similar ones seen at 6\,cm and 11\,cm. Such different shapes toward higher frequencies 
were also seen previously in this source (Fuhrmann et al. 2008) and were interpreted as 
indication for the presence of physically different variability mechanisms, as e.g.
interstellar scintillation dominating at the longer wavelength and a with frequency
increasing contribution of source intrinsic variability. In this context, 
variability time scales which are seen in the 2.8\,cm data, 
but are not seen in the 11\,cm and 6\,cm data, may be related to some
source intrinsic variability and may well therefore appear also in the optical, 
which is not affected by interstellar scintillation.

In the optical bands S5~0716$+$714 also displayed remarkable variability during the campaign period. 
The most dense coverage at optical frequencies was obtained in the V and R passbands. 
The SFs of these data are shown in  Fig.\ \ref{fig3}. They show rapid variability 
with multiple cycles of rises and declines. Differences between both SFs are likely due to
the different time-sampling in both data sets and the location of observing gaps within the 
two data trains.
The B and I passband light curves are much sparser and so cannot reveal significant 
features in the SF analysis. We therefore focus on the SFs for R- and V-bands.
The first break in both R and V band SF curves 
appears at a timescale of $t_{min} = 0.25\pm0.05$ days. At R-band the SF 
shows pronounced maxima also at $0.50\pm0.05$ days, $0.90\pm0.05$, 
$1.4\pm 0.1$ days and $1.9\pm 0.1$ days.
At V-band the variations are less rapid, with a dominant
time scale at $0.8\pm0.05$ days and $1.9\pm 0.1$ days.
The 0.5\,day and the $1.4\pm 0.1$ day time scale, seen in R-band, are not seen 
in V-band.

We note that these time scales are within their measurement 
errors factor of two multiples of each other, providing a hint of  
an underlying 0.25\,day quasi-periodicity (QPO). We further note that 
at 2.8\,cm the shortest timescale seen in the SF is 0.5\,days, which is 
very close to the second harmonic of the optical variability.
This might indicate some correlation
between radio and optical. We check the possibility for such a correlation in
Section 3.3. We investigate the possibility for QPOs being present in the
optical data using other techniques in Section 3.6.

\subsection{Auto- and Cross-correlations}
To quantify the correlation among the multi-frequency light curves of the source during the campaign
period, we computed the discrete auto and cross correlation functions (ACF, DCF)
between different frequencies to search for possible time lags. We followed Edelson $\&$ Krolik (1988)
and Lehar et al.\ (1992) to calculate the DCF with details given
in Rani et al.\ (2009).

In Figure \ref{fig4} we plot the two auto-correlation functions (ACF) of the 6\,cm data from the Effelsberg
and Urumqi telescopes. The differences seen in the shape of the two ACFs is explained by the different
data-quality from the two telescopes, with a higher signal-to-noise ratio and smaller beam size
(and therefore lower in beam confusion) at the larger telescope.

In Figure \ref{fig5} we show the DCF of the 6\,cm data from the Effelsberg
and Urumqi telescopes. 0716+714 was observed at both telescopes with very similar time sampling and
time coverage. The high degree of correlation of the two data trains is obvious and convincingly
demonstrate the reality of the observed intra-day variability.
No significant time lag ($\tau_{6.0/6.0} = 0.0\pm0.05$) is seen between the two telescopes.

\begin{figure}
\epsfig{figure = 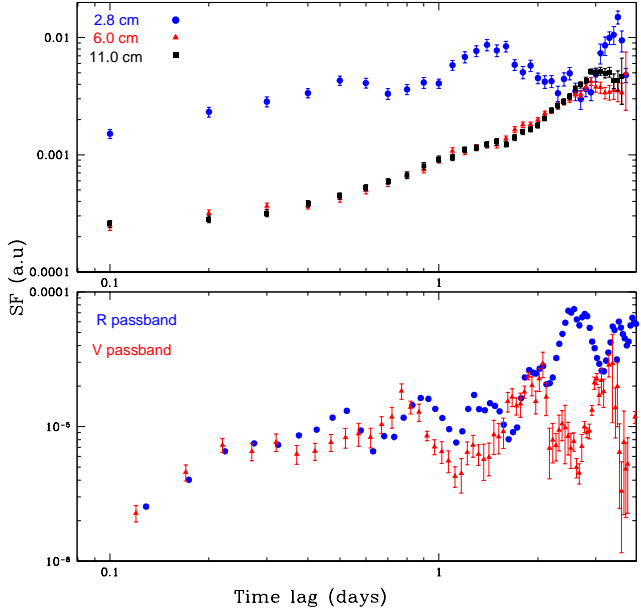, scale=0.43}
\caption{Structure function analysis curves of S5 0716+714, with radio in the upper panel and
optical in the lower. Different symbols are 
used for different wavelengths. }
\label{fig3}
\vspace*{0.5cm}
\end{figure}

\begin{figure}
\centerline{\includegraphics[scale=0.35]{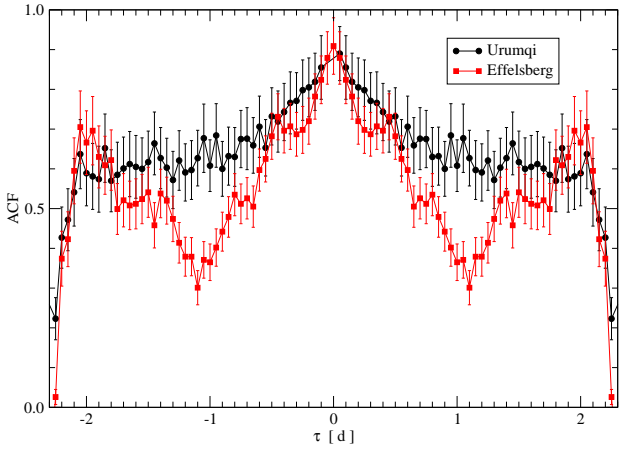}}
\caption{Auto-correlation functions of the 6.0\,cm data from Effelsberg
and Urumqi. }
\vspace*{0.5cm}
\label{fig4}
\end{figure}

\begin{figure}
\centerline{\includegraphics[scale=0.35]{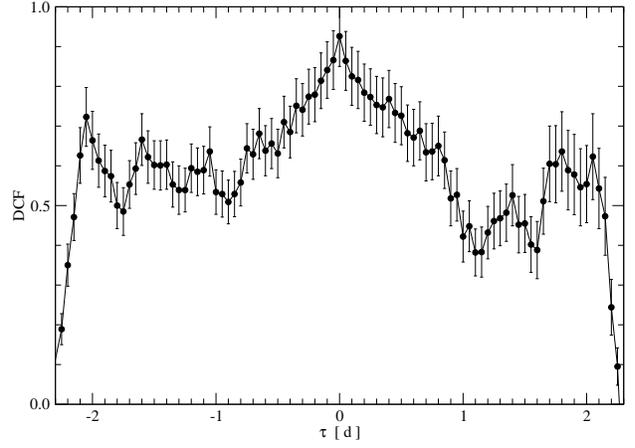}}
\caption{Cross-correlation function of the 6.0\,cm data between the Effelsberg and the Urumqi
telescopes. }
\vspace*{0.5cm}
\label{fig5}
\end{figure}

\begin{figure}
\centerline{\includegraphics[scale=0.35]{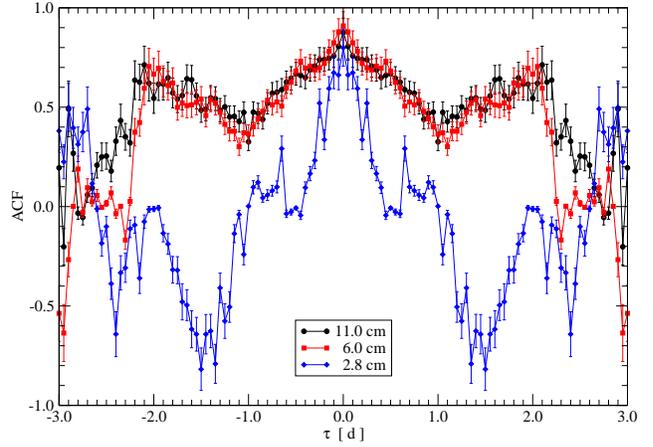}}
\caption{Auto-correlation functions of the 11.0\,cm (circles),
6.0\,cm (squares) and 2.8\,cm (diamonds) data from Effelsberg. }
\vspace*{0.5cm}
\label{fig6}
\end{figure}

Similar to the structure function analysis (see Fig. \ref{fig3}) the auto-correlation functions
of the 11\,cm and 6\,cm Effelsberg data show a common shape and show similar decorrelation time-scales
(Fig.\ \ref{fig6}). On the other hand, the 2.8\,cm data decorrelate much faster, indicative of
faster variability at this shorter wavelength (higher frequency). The ACFs of the 11\,cm and 6\,cm
bands also suggest a regular pattern on a $\sim 2$\,day timescale. Since this timescale is
just half of the duration of the overall $\sim 4$\,day time coverage of the radio data,
we do not regard this timescale as significant.

\begin{figure}
\centerline{\includegraphics[scale=0.35]{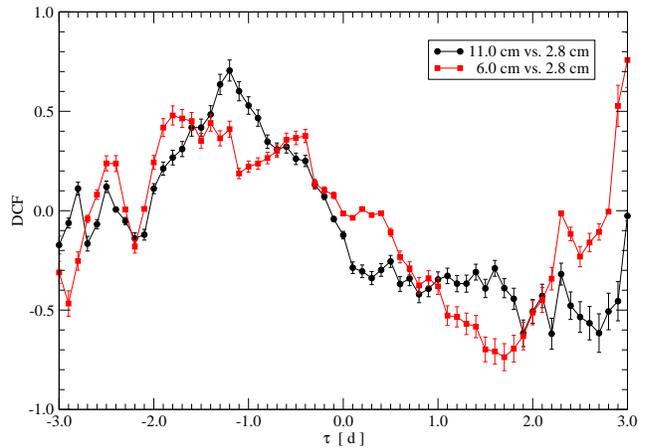}}
\caption{Cross-correlation functions of the Effelsberg data: circles denote 11.0\,cm versus 2.8\,cm,
and squares denote 6.0\,cm versus 2.8\,cm. }
\vspace*{0.5cm}
\label{fig7}
\end{figure}

To search for possible timelags between the three radio bands, we performed a cross-correlation
analysis of the Effelsberg radio data. Figure \ref{fig7} shows the two independent cross-correlations
of the 11.0\,cm and 6.0\,cm light curves versus the 2.8\,cm data. Formally we calculate
a time lag of $\tau = - (1.2 \pm 0.1)$\,day between 11.0\,cm and 2.8\,cm in the sense that the 2.8\,cm
data lead. Such frequency dependence is consistent with the canonical behavior seen in AGN
(e.g.\ van der Laan 1966, Marscher \& Gear 1985).

\begin{figure}
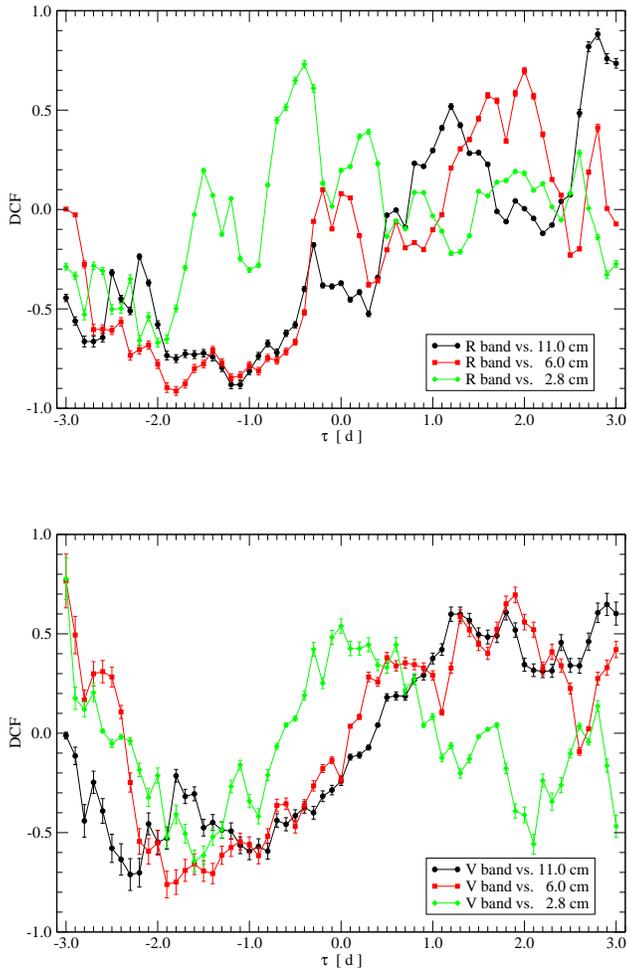

\epsfig{figure = fig8a.eps, scale=0.35}
~~\\
~~\\
~~\\
~~\\
\epsfig{figure = fig8b.eps, scale=0.35}
\caption{The cross-correlation analysis curves of optical R passband light 
curve (top) and optical V passband light curve (bottom)
with all the three radio wavelength light curves of the source. 
A bin size of 0.1 day is used in the correlation analysis.  }
\vspace*{0.5cm}
\label{fig8}
\end{figure}

\begin{figure}
   \centering
\epsfig{figure = 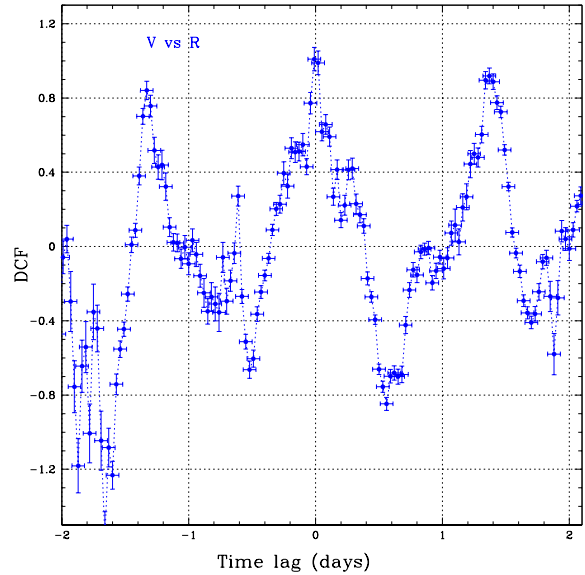, scale=0.4}
\caption{Cross-correlation analysis curves between the optical V- and R-band light
curves.  Although a bin size of 0.01 day is used in the correlation analysis this curve is 
shown with a bin size of 0.02 day.  }
\label{fig9}
\end{figure}

\begin{figure}
\epsfig{figure = 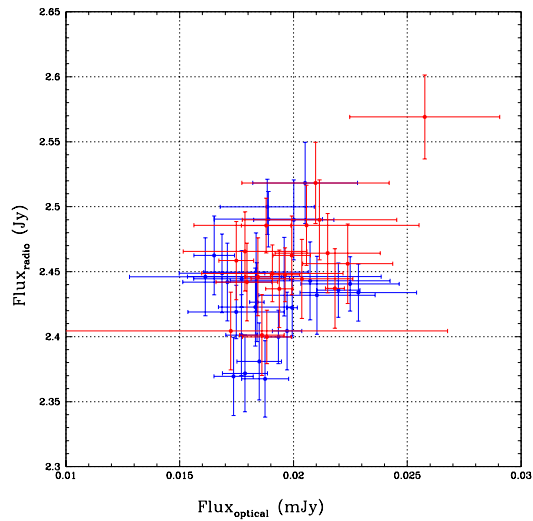, scale=0.35}
\caption{The 2.8\,cm radio flux plotted versus optical R-band flux.
The blue symbols show the original data, the
time shifted 2.8\,cm radio data are shown in red (a 0.35 day shift was applied, see text).  }
\vspace*{0.5cm}
\label{fig10}
\end{figure}

\begin{figure}
\epsfig{figure = 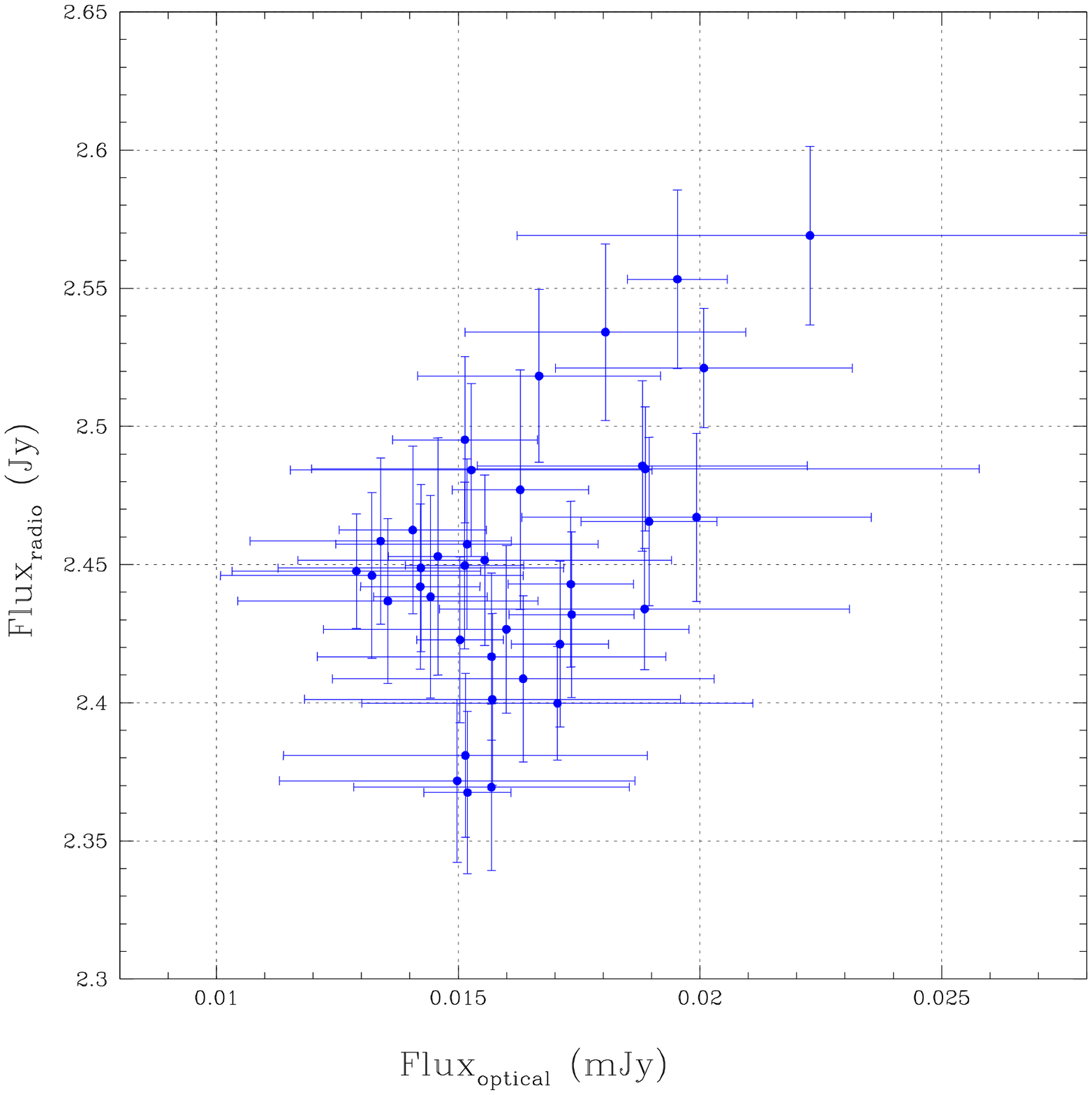, scale=0.35}
\caption{The 2.8\,cm radio flux plotted versus optical V-band flux.  }
\vspace*{0.5cm}
\label{fig11}
\end{figure}

Next, we look for a possible correlation with flux variations at optical-radio frequencies. 
Figure \ref{fig8} (top) may indicate a correlation with $\sim (0.35 \pm 0.15)$\,day time lag
between R and and 2.8\,cm, in the sense that radio is leading over optical, which apparently
is opposite to physical expectations (the higher frequency should come first). 
At V-band the maximum correlation with 2.8\,cm occurs at zero time lag, 
however, with a lower cross-correlation coefficient than for 
R-band (DCF peak at: 0.73 for R-band, 0.53 for V-band). 
Since the R- and V-bands cross-correlate very strongly with zero time lag (Fig.\ \ref{fig9}), we think that 
the above time lag of $\sim 0.35$\,day is unlikely to reflect physical reality.   
The R-band light curve shown in Fig.\ref{fig1} is less continuous and shows larger 
time gaps, while the V-band curve is more continuous and therefore better suited for
the search of a possible cross-correlation. 
We generate flux-flux plots to quantify a possible correlation between optical and 2.8\,cm data.
In Figure \ref{fig10} we show the results using R-band data and in Figure \ref{fig11}
with V-band data. 
Formally we obtain the following correlation coefficients ($r$ being the linear Pearson correlation coefficient)
and significances. From Figure \ref{fig10}: 
R versus 2.8\,cm: $r$ =0.163 and 58.59\, per cent significance;
R versus time shifted 2.8\,cm: $r$= 0.687 and 99.90\, per cent significance.
From Figure \ref{fig11}: 
V versus 2.8\,cm: $r$=0.52 and 99.92\, per cent significance.
However, if we remove the highest data point in Figure \ref{fig10} \& \ref{fig11}, the 
correlation significance drops below 95 per cent for R-band versus 2.8\,cm while it is still sustained 
for V versus 2.8\,cm. 

We therefore conclude that we cannot claim a significant detection of a correlation between 
optical R-band and 2.8\,cm radio band, probably due to the limited time sampling of
the R-band data. In the more continuous V-band, however, such a correlation may exist,
although we regard the evidence for it as weak, and the nominal time lag as physically unlikely.

\subsection{Spectral energy distribution} 

From the available multi-frequency data of this campaign we can construct a 
quasi-simultaneous SED of the BL Lac S5~0716$+$714. For the SED we averaged in time all flux 
density measurements obtained at the different observatories during the period December 11-15, 2009.
For radio through X-ray bands the length of the error bar reflects the strength of the variability.
At GeV frequencies the error bars are derived from 
the spectral fitting. As the IDV observations were carried out only at three 
radio wavelengths, we also included additional flux density measurements 
at other wavelength from the IRAM 30-m Pico Veleta telescope and some public resources. 
These measurements were obtained quasi-simultaneously during the period 
December 9--16, 2009 and will be discussed in more detail in a forthcoming paper
on the long term variability of 0716+714 (Rani et al.\ 2012).

\begin{figure}
   \centering
\epsfig{figure = 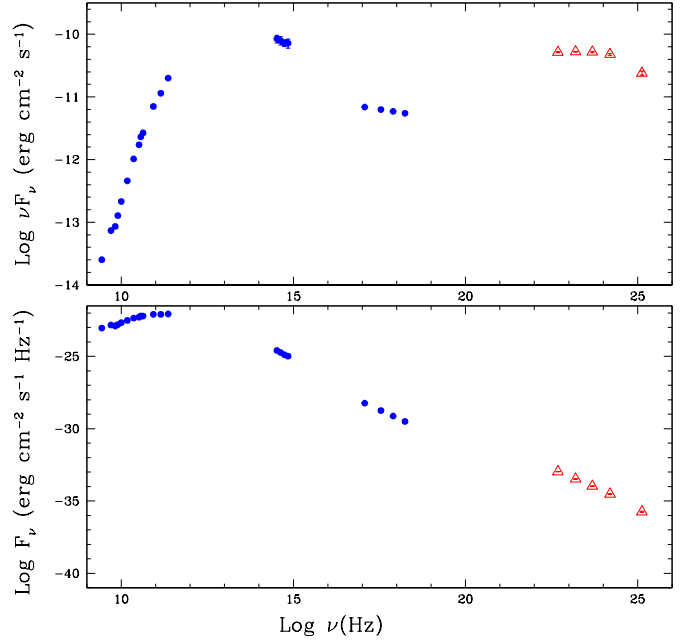, scale=0.45}
\caption{Broad band (radio to $\gamma$-ray) spectrum of S5~0716$+$714 over the campaign period.
Simultaneous data are in blue color. For comparison we also include the non-simultaneous 
$\gamma-$ray data from Fermi/LAT from the 2FGL catalogue (red symbols, see text).
The top panel shows the spectral energy ($\nu F_\nu$), the bottom panel flux density ($F_\nu$)
plotted versus frequency. For the radio through X-ray bands the error bars represent here 
the variations in flux density rather than the measurement uncertainty.}
\label{fig12}
\end{figure}

The SED of the source is displayed in Fig.\ \ref{fig12}. The upper panel displays the $\nu$ - $\nu F_{\nu}$ plot, while the  $\nu$ - $F_{\nu}$ plot is shown in lower panel. The 
maximum luminosity of the SED appears to be in the range of around 10$^{12}$ -- 10$^{14}$ Hz,
which, unfortunately is in a spectral region with no data.

In the SED we also included the public available $\gamma-$ray data from the
second AGN catalogue of the Fermi satellite (2FGL catalogue; Ackermann et al.\ 2011). 
These data result from a 2 year average (Aug. 2008 - Aug. 2010) of the source flux density and therefore 
are not truly simultaneous to our observing campaign. We note however that 0716+714 
was in a low $\gamma-$ray state at the time of our observation, so that the expected GeV flux
in December 2009 are likely be lower than the 2FGL values plotted in this SED. 

\subsection{Jet Doppler factors}

Within the framework of synchrotron self-Compton models it is  possible to 
constrain the Inverse Compton Doppler factor ($\delta_{IC}$)
by comparing the expected and observed fluxes at 
higher frequencies (Marscher 1987 and Ghisellini et al. 1993). 
This IC Doppler factor is defined as  
\begin{eqnarray}
\delta_{IC} \geq f(\alpha)S_{m}\left(\frac{\ln(\nu_{b}/\nu_{m})\nu_{\gamma}^{\alpha}}{S_{IC}\theta_{\nu}^{(6-4\alpha)}\nu_{m}^{(5-3\alpha)}}\right)^{1/(4-2\alpha)}(1+z)\,\,
\label{doppler_IC}
\end{eqnarray}
where $\nu_{b}$ is the synchrotron high frequency cut-off in GHz, $S_{m}$ the flux 
density in Jy at the synchrotron turnover frequency $\nu_{m}$, $S_{IC}$  the observed 
$\gamma$-ray flux in Jy (assumed to arise from the IC process) at $\nu_{\gamma}$ in keV, $\alpha$ is the spectral index of the optically thin part of the spectrum, $\theta_{\nu}$ the angular source size in mas 
and $f(\alpha)\simeq 0.14-0.08\alpha$. For the high energy cut-off we follow
Fuhrmann et al.\ (2008) and use $\nu_{b} \sim 5.5 \times 10^5$ GHz. 
The angular size of the rapidly varying region $\theta_{\nu}$ is constrained to be $\leq$0.03 mas 
as used by Agudo et al.\ (2006) and consistent with size measurements from mm-VLBI. 
We adopt this value in our further calculations. For the radio spectrum
we use an optically thin spectral index of $\alpha = -0.44$, 
and for the synchrotron turnover we adopt $S_{m}$ = 8.9 Jy and 
$\nu_{m}$ = 148 GHz (for details of spectral fitting see Rani et al.\ 2012).

The estimated limits on $\delta_{IC}$ are 18.8 (adopting $S_{IC}$  = 1.77$\times 10^{-6}$ Jy at 
$\nu_{\gamma}$ = 1.4 keV) and $\delta_{IC} =26.1$ (adopting $S_{IC}$  = 1.73$\times 10^{-7}$ Jy at 
$\nu_{\gamma}$ = 7.2 keV).  If the same calculation is applied for the GeV fluxes and
if we assume that the same photons would be producing also the gamma-rays, 
then a considerably higher Doppler factor of $\delta_{IC} = 69$ is obtained 
(adopting $S_{IC} = S_{\gamma}$ = 1.95$\times 10^{-11}$ Jy at 
$\nu_{\gamma}$ = 5.47$\times 10^{5}$ KeV)\footnote{The GeV fluxes were obtained 
from public available FERMI/LAT light curves at 
http://fermi.gsfc.nasa.gov/ssc/data/access/lat/msl$\_$lc/}.
For a given Lorentz-factor $\gamma$, the maximum Doppler-factor is $\sim 2 \gamma$.
A Doppler-factor of up to $\sim 50$ therefore is consistent with the observed jet
speed, which is in the range of $\gamma$ = 15--25 (Bach et al.\ 2005, Fuhrmann et al.\ 2008,
Rastorgueva et al.\ 2009). 

In the inverse Compton process it is usually assumed that the synchrotron photons
from the IR-optical branch are upscattered up to GeV energies.
If we assume that the jet is not stratified, i.e. that the 
Doppler-boosting is homogeneous over the emission region, we find that the observed GeV flux 
could also be explained via inverse Compton scattering from a synchrotron component with a
spectral maximum in the 300-400 GHz regime. The possibility of SSC
dominance in Compton dominated blazars has been investigated in detail in Zacharias
\& Schlickeiser (2012).   

It is also possible to obtain a limit on the Doppler factor $\delta$ by assuming that the 
high-energy $\gamma$-ray photons can collide with the softer radiation to produce e$^{\pm}$
pairs. The cross-section of this process is maximized at $\sim \sigma_{T}/5$ (see 
Svensson 1987 for details), where $\sigma_{T}$ is the Thompson scattering 
cross-section. This leads to  a lower limit on $\delta$ (following Dondi \& Ghisellini 1995):
\begin{equation}
\delta \geq \Big [ 3.5 \times 10^3 \frac{(1+z)^{2\alpha} (1+z-\sqrt{1+z})^2 F_x (3.8 \nu \nu_x)^{\alpha}}{t_{var}} \Big ]^{1/(4+2\alpha)}
\end{equation}  
where $F_x$ is the X-ray flux in $\mu$Jy and $\nu_x$ is the corresponding X-ray frequency in keV, 
$\nu$ is the GeV frequency, $\alpha$ is the spectral index measured between 1 KeV and 100 MeV. Using
$F_x$ = 1.77 $\mu$Jy, $\nu_x$=1.44 KeV, $\alpha=-$1.05, we obtained $\delta \geq 24$ for 300 MeV and 
$\delta \geq 12$ for 1 GeV energies. Since the published GeV data are not simultaneous with our
radio to X-ray data, the accuracy to which the Doppler factors can be estimated is limited. 
A more accurate, though not largely different, estimate of the Doppler boosting will be presented in a 
forthcoming paper, which is based on  truly simultaneous gamma-ray data (Rani et al. 2012).

From all of the above approaches we think it reasonable to conclude 
that the blazar 0716+714 exhibits a 
high Doppler factor of at least 12 and most likely  $\geq 20-25$,
which is not typical for a BL\,Lac type object.

\subsection{Nearly periodic variations in the optical?}

\begin{figure}
\centerline{\includegraphics[width=4.0in,height=7.0in,angle=-90]{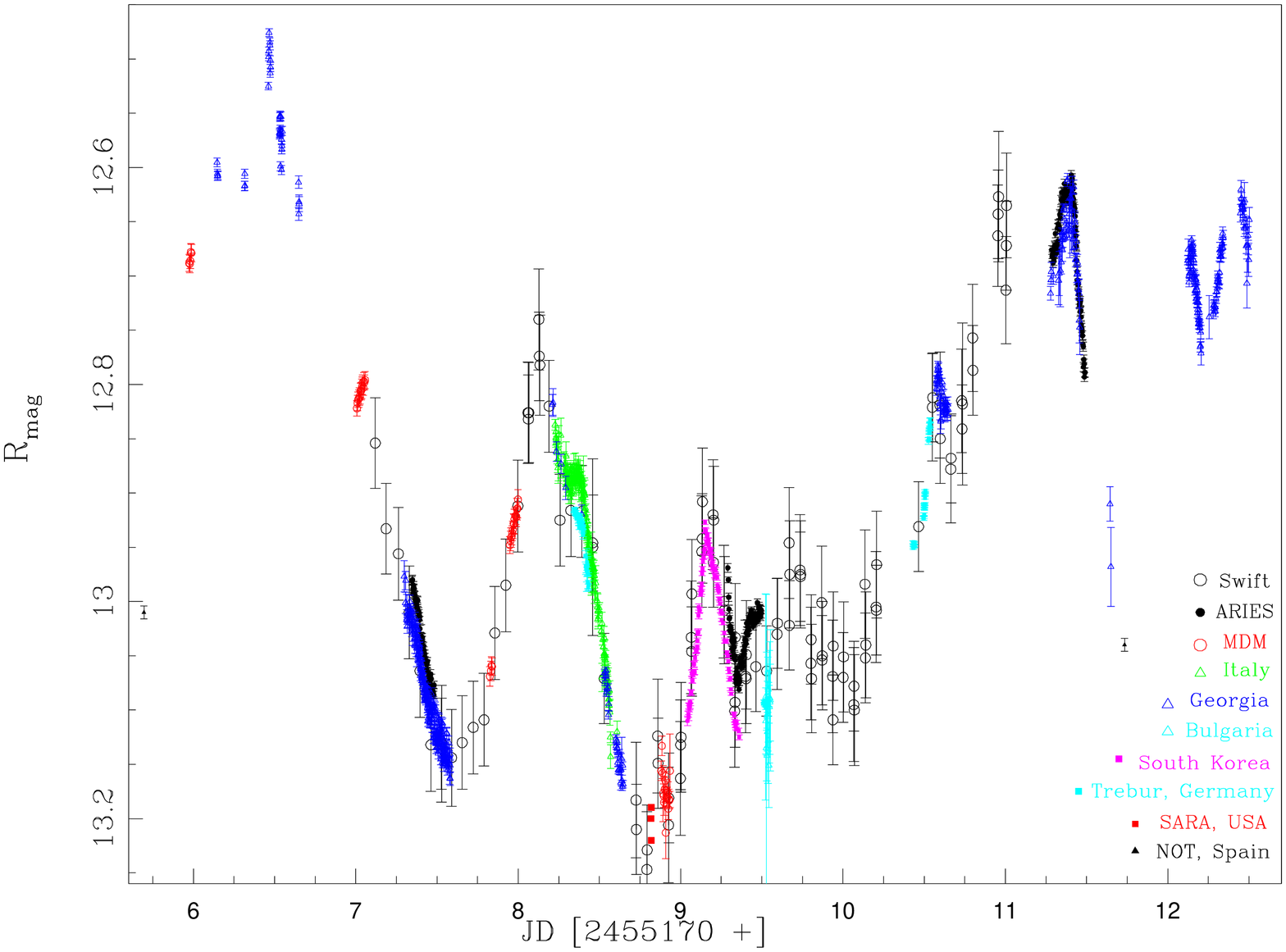}}
\caption{R passband light curve of the blazar S5~0716$+$714. Different symbols and colors denote
for data from different observatories. Color corrected V-band data from Swift 
were added to complement the time coverage.}
\label{fig13}
\end{figure}

The complete optical light curve is given in Figure \ref{fig13}, 
with the different observatories labelled.  
It is important to note the excellent agreement between 
the data taken at different telescopes during the many periods when we had essentially 
simultaneous coverage. Over the course of the 7 total days of these observations the R magnitude of S5~0716$+$714 varied between 12.48 mag and 13.25 mag.
Strong variations were seen throughout the campaign, with frequent excursions of 
0.2 -- 0.4 mag over a few hours. Such changes are not uncommon for this active 
blazar (e.g. Montagni et al.\ 2006; Gupta et al.\ 2008; Rani et al.\ 2011 and 
references therein).  The duty cycle in previous observations was found to be 
unity, with some detectable variation every time it is observed for more than 
a few hours (Wagner \& Witzel 1995, Wagner et al.\ 1996) and our results 
certainly agree with those earlier data.  We have quantified these variations by examining 
them for possible timescales and quasi-periodic fluctuations.

Following Mohan et al.\ (2012), we employ a suite of statistical techniques to analyze the optical 
light curve (Fig. \ref{fig11}), namely, Fourier periodogram, wavelet analysis, 
multi-harmonic analysis of variance (MHAoV) and Lomb-Scargle periodogram (LSP).  
The LSP and MHAoV offer a natural way of period detection when the data is 
unevenly sampled with the MHAoV also able to detect non-sinusoidal signals.  
Together, these are designed to detect possible QPOs and test the 
validity of a detection using Monte-Carlo simulations based significance 
testing.  This suite helps in making QPO detection reliable through consistent detection 
with significance testing in all techniques. In addition, the phase of existence and 
number of cycles a QPO is present during the entire length of the observations is 
obtainable from a wavelet analysis. As an example of this suite technique 
we apply it to our new data on 0716+714.

\begin{figure}
\centerline{\includegraphics[scale=0.5]{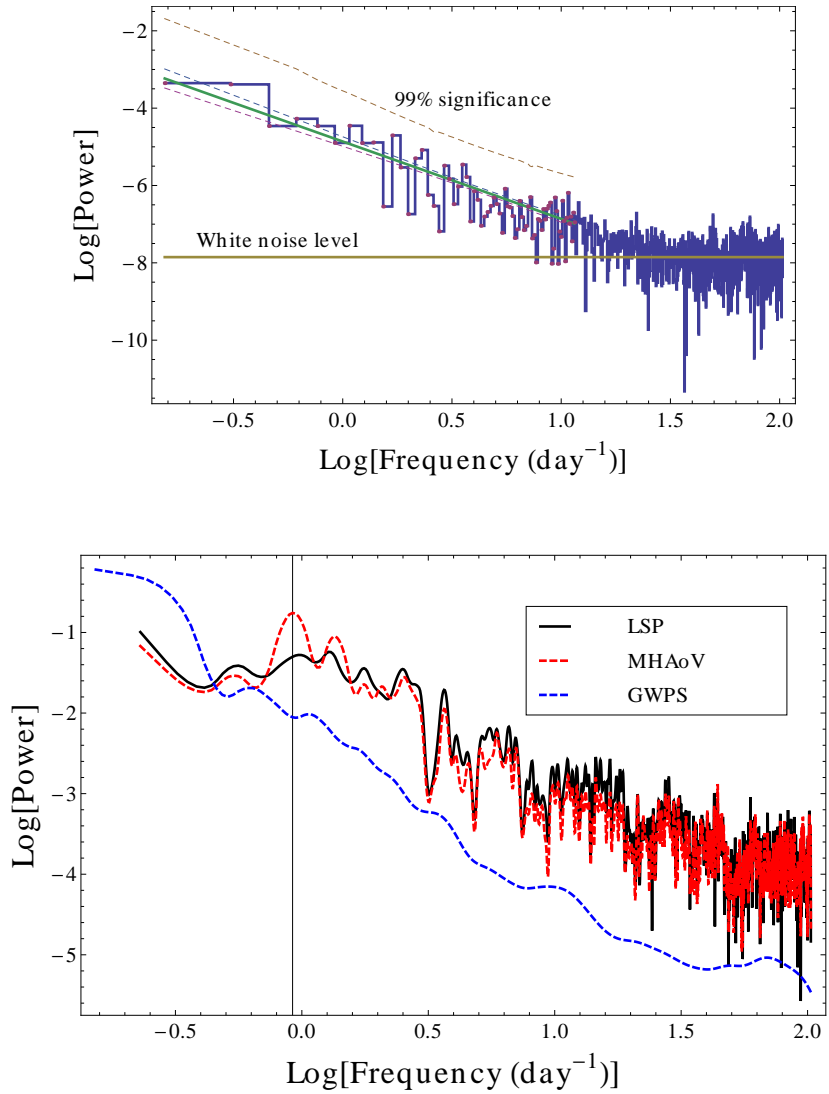}}
\caption{Top plot: periodogram on a log-log scale. The yellow solid line indicates
the white noise level which is the contribution of photon shot noise at
all
frequencies. The green solid line indicates the best fit power law to the
log-log periodogram. The dashed lines above and below the best fit line
indicate the errors on the power law model. The upper yellow dashed line
indicates the 99 per cent significance contour, taking into account errors
on the power law as well as model uncertainties. Bottom plot: Combined
plot showing the LSP, MHAoV and the GWPS on a log-log scale with a black
vertical line at the position of the periodicity of 1.1 day. The MHAoV and
LSP indicate a feature at 1.1 days with significances of 94.6 \% and 80 \%
respectively. The GWPS shows a minor feature at 0.93 days just to the
right of this. All the curves are normalized such that the total power
summed over all frequencies between the sampling and the Nyquist frequency
are equalized.}
\label{fig14}
\end{figure}

\begin{figure}
\centerline{\includegraphics[scale=0.5]{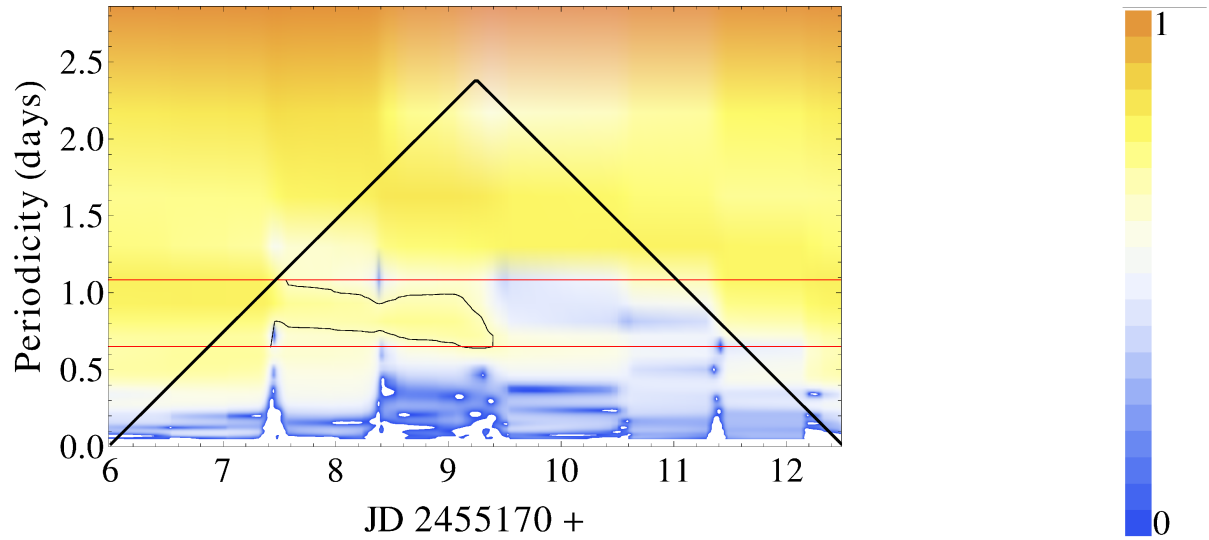}}
\caption{Wavelet contour plot beside which is an intensity scale indicating the color scaling of the wavelet power. Contour plot abscissa: time duration of observation; ordinate: periodicities searched for in the detection of quasi-periodic signal. A broad feature between JD 2455170 + 7.4 and 9.4 days is indicated inside a contour drawn between periodicities of 0.65 days and 1.08 days (horizontal red lines) within the cone of influence. This contour indicates the wavelet power above 90\% confidence in this region. A periodicity of 0.93$^{+0.15}_{-0.28}$ is indicated which lasts for 2.2 cycles.}
\label{fig15}
\end{figure}

The fitting of the Fourier periodogram (top plot in Fig. \ref{fig14}) follows the procedure 
described in Vaughan (2005). It is well fit by a power law model with a slope of 
$-2.0\pm0.16)$. 
The Kolmogorov-Smirnov goodness of fit test yields a $p$ value of 
0.99 indicating a good fit. Errors on the fit parameters, the slope and normalization 
are determined. Then, the error on the power law model is evaluated, after which model 
uncertainties are accounted for before placing a 99\, per cent significance contour 
(top plot in Fig. \ref{fig14}). There are no significant detections above this level. 
Monte-Carlo (MC) simulations are carried out using the Timmer \& K{\"o}nig (1995) 
algorithm to generate 6600 random light curves with a range of slopes within the fit 
parameter errors to simulate the properties of the original light curve. 
The MHAoV and LSP are determined for each of the generated light curves and an estimate of 
the number of times the data based MHAoV and LSP ordinates at the interesting periodicity 
are above the simulated ordinates gives a measure of the significance. The MHAoV (bottom 
plot in Fig. \ref{fig14}) detects a possible period at 1.1\,days with a probability of 94.6\, per cent
from the MC simulations based significance test. The LSP (bottom plot in Fig. \ref{fig14}) 
detects the same period of 1.1 days with a probability of 80\, per cent from the MC simulations 
based significance test.
The wavelet analysis (Fig. \ref{fig15}) indicates a quasi-period of 0.93$^{+0.15}_{-0.28}$ days in the duration between $\sim$ 7.4 days and 9.4 days within the cone of influence (triangular bounded region in between the two black lines in Fig. \ref{fig15}) indicating that it lasts for 2.2 cycles.
By integration of the wavelet signal along the time axis (abscissa in Fig. \ref{fig15}), the global wavelet power spectrum (GWPS) is obtained (bottom plot of Fig. \ref{fig14}).

Since the nominal periodicity shows a fairly large range of 0.65 days - 1.1 days 
in the optical data, and since it lasts only for 2 -- 3 cycles, any periodicity 
that may be present is certainly broad and weak.
We can argue that a timescale of $\sim 1$ day has been seen in the optical variations during 
our campaign as it is supported by the moderate significance estimated for the MHAoV and LSP detections.
However, despite having a relatively large amount of optical data from telescopes at different longitudes, 
there were gaps during the campaign that were only filled in by the {\it Swift} UVOT monitor.
These data are of lower accuracy than the bulk of the ground-based data and they affect the sensitivity 
of our tests for QPOs.

\section{Discussion and Conclusions}

We have successfully carried out a multi-wavelength campaign to observe the well-known
blazar S5\,0716$+$714. The optical portion of the data that we have reported here was obtained
from 9 ground-based and one space-based telescopes over almost 7 days.  The target was
very active, with repeated `mini-flares' of 0.2--0.7 mag in amplitude and an overall 
variability of 0.8 mag in the optical V-band.
Significant peak-to-peak variations of the order of 5-10\, per cent were also seen in the radio 
bands but only marginal variability was detected in the X-rays. In the radio bands the variability
amplitudes increase with frequency, and the variability time scales become shorter. 
We find significant correlations between the different radio bands as well as between 
the optical passbands, but no significant correlations between radio/optical and X-ray bands.

The visual inspection of the optical light curve suggests the presence
of a characteristic variability timescale. Therefore  
we analyzed this formally using the periodogram, LSP, MHAoV and wavelet techniques. 
A timescale in the range of $0.9 - 1.1$ days is determined from these.
Because of its moderate formal significance (80 per cent from LSP and 94.5 per cent from MHAoV analyses) 
and a small number of putative cycles (2.2 seen in the wavelet analysis) a formal claim of
strict (or quasi-) periodicity cannot be made. 

The structure function analysis yields for the three fastest variability modes
timescales of 0.25, 0.5 and 1.0 day in the optical bands. In the radio bands
the three fastest variability modes appear at 0.5, 1.1 and 1.5\,days. Within the
measurements' uncertainties, the variability timescales at 0.5 day and 1.0 day 
appear common between the radio and optical. It may be noteworthy that these timescale 
are essentially integer multiples of the fastest optical variability timescale of 0.25\, days.
It is unclear whether this is a chance coincidence or reflects
a common physical origin.

A comparison of the observed modulation indices at radio bands with the expected
strength for (weak) interstellar scintillation (ISS) suggests that most of the observed radio IDV is 
in agreement with a ISS slab model, which is able to reproduce the observed variability
time scale of the order of $\sim 1-2$\,days. At 2.8 cm, however, the observed variability index is 
larger than expected from the model which could be due to some `underlying' source intrinsic variability. 
This interpretation is supported by the detection of a time lag between 2.8\,cm and the longer wavelengths
in the sense that the 2.8\,cm variability is leading. This is typical for AGN variability
and is commonly interpreted as a source intrinsic opacity effect.

We note a possible cross-correlation between 2.8\,cm and optical V-band.
Although mathematically significant, such a correlation must be regarded with 
some skepticism in view of the lack of a similar correlation at R-band, 
the unexpected sign of the putative time lag (radio is leading) and the
limited length of the data trains.
So, we cannot claim any significant correlation between the flux variations at 
optical and radio frequencies over this 1-week IDV campaign. We further
note that in their analysis of the long-term variability of 0716+714 
during 2007-2011, in which 0716+714 shows a prominent flare with peak in late December 2009,
Rani et al.\ (2012) find that the optical flux variations lead the radio by $\sim (60-70)$ days. 
Such a time lag 
would exclude any direct radio-optical correlation
in the 1 week long data train discussed here.

As a matter of fact, correlated radio-optical variability of this source has been 
investigated in several earlier observing campaigns. During four weeks of 
multi-frequency observations in February 1990, Quirrenbach et al.\ (1991) detected 
a significant one-to-one correlation between optical/radio flux variability
(see also Wagner et al. 1996, Qian et al. 1996). Such a correlation was not seen in 
two later radio-optical IDV campaigns, which were performed
in 2000 (unpublished data) and in 2003 (Ostorero et al.\ 2006, Agudo et al.\ 2006, 
Fuhrmann et al.\ 2008).

To further explore the possibility of a radio-optical correlation,
we examine the spectral characteristics of 0716+714 for three different 
radio-/optical IDV campaigns. In Table \ref{tab_camp} we summarize
for each observing date the spectral properties of 0716+714, where we used
flux densities averaged over the duration of the campaign.
In the table we list the synchrotron peak frequency ($\nu_{max}$) in column 2, 
the radio spectral index in column 3, and the mm-optical spectral index in column 4.
We see that in 1990 the source showed a factor $\sim 3$ 
lower synchrotron turnover and considerably steeper radio spectrum than in the two later campaigns,
where 0716+714 was much more active and showed prominent flux density outbursts.

It is therefore likely that the detection/non-detection of correlated radio-optical 
IDV relates to the activity state of the source.  IN particular, the opacity of the 
emitting region in the cm-bands (e.g.\ Qian 2008) and the presence of high or 
low peaking flux density flares, which trace moving shocks in a relativistic 
jet at different separations from the jet base (e.g.\ Valtaoja et al.\ 1992), may determine 
the presence of such correlations. Near and above 
the actual turnover frequency, the opacity decreases progressively, the spectrum
steepens and the time lag between radio and optical variations vanishes. 
The direct radio-optical correlation seen in February 1990 implied
a small ($< 1$ day) time lag in the variability, which is fully 
consistent with the observed optically thin state and a time 
of intermittent quiescence in the overall source variability.

\begin{table}
\caption{ A comparison of spectral variations over different campaign epochs  }
\begin{tabular}{lcccc}
Epoch       & $\nu_{max}$    & $\alpha_{radio}$  & $\alpha_{mm-optical}$ & Reference   \\\hline
Feb. 1990   & 35$\pm$5        & -0.82$\pm$0.08            & -0.75$\pm$0.05    & 1   \\
Nov. 2003   & 90$\pm$5        & -0.35$\pm$0.10            & -0.88$\pm$0.04    & 2   \\
Dec. 2009   & 114$\pm$12      & -0.44$\pm$0.10            & -0.85$\pm$0.03    & 3   \\\hline
\end{tabular}  \\
$\nu_{max}$ : synchrotron peak frequency in GHz from spectral fit \\
$\alpha_{radio}$ : optically thin radio spectral index calculated by fitting a 
synchrotron self-absorbed model over a frequency range 2.7 - 230 GHz.  \\
$\alpha_{mm-optical}$ : spectral index from 230 GHz to optical R-band. \\
References : 1. Quirrenbach et al. 1991; 2. Ostorero et al. (2006) and 3. this paper . 
\label{tab_camp}
\end{table}  

The detection of QPOs with similar time scales in radio and optical band
would restrict their physical origin to jet rather than to accretion disk 
physics. As the activity state of the source and its time variable opacity play a 
fundamental role in the search for correlated radio-optical IDV,
we see two ways to proceed in the near future:
(i) observe during times of low activity, e.g.. minimal flux and steep spectrum,
and/or (ii) observe at frequencies above the critical turnover frequency,
which for 0716+714 is at $\geq 100$\,GHz (in mm- and sub-mm bands).
In the optically thin regime the lower source opacity
will ensure that the observer actually looks at the same physical region. In order
to facilitate the identification of QPOs, data trains have to be long enough
in time so that a sufficiently large number of variability cycles can be observed. 
With a typical variability time scale of 0.5-1.5 days in 0716+714, at least $5-10$ 
duty cycles are necessary to unambiguously identify a QPO. This means that a continuous 
broad band monitoring for more than 1 week duration is necessary. In order to
separate the intrinsic variability from unavoidable interstellar scintillation (caused
by the small size of the emission region) 
multi-frequency observations are required, with a frequency coverage which should 
include the millimeter and sub-millimeter radio bands. This will help
to avoid the effects from interstellar scintillation, which dominate at the longer
cm-wavelengths.

\section*{Acknowledgments}

We thank the referee for constructive comments that have helped us to improve
the paper.
We acknowledge data from the 100\,m Effelsberg radio telescope, which is operated by
the Max-Planck-Institute f\"ur Radioastronomie in Bonn (Germany).
This paper made use of the data obtained with the Urumqi 25m radio telescope of Xinjiang 
Astronomical Observatory of the Chinese Academy of Sciences (CAS).
The IRAM 30m Telescope is supported by INSU/CNRS (France), MPG (Germany), and IGN (Spain).
We thank the {\it Swift} team for making these observations possible. 
The work of ACG, BR, RB, HG, SP, ES and AS was partially 
supported by Scientific Research Fund of the Bulgarian Ministry 
of Education and Sciences (BIn-13/09 and DO 02-85) and by 
Indo Bulgaria bilateral scientific exchange project INT/Bulgaria/B-5/08 funded by DST, India.  
The work of PJW is partially supported by US NASA grant NNX11AB90G.    
The   Abastumani Observatory team acknowledges financial support by the
Georgian National Science Foundation through grant GNSF/ST09/521$\_$4-320. 
The work of K\'EG was partially supported by the COST Action
MP0905 ``Black Holes in a Violent Universe" and by the Hungarian
Scientific Research Fund (OTKA, grant no. K72515).
IA acknowledges the funding support by the regional government of Andaluc\'{i}a, and by the Ministerio de Ciencia e Innovaci\'{o}n of Spain through grants  P09-FQM-4784 and AYA2010-14844, respectively.
TP acknowledges data based on observations made with the Nordic Optical Telescope, operated
on the island of La Palma jointly by Denmark, Finland, Iceland,
Norway, and Sweden, in the Spanish Observatorio del Roque de los
Muchachos of the Instituto de Astrofisica de Canarias.

{}

\clearpage

\onecolumn

\end{document}